\documentclass[12pt]{iopart}

\expandafter\let\csname equation*\endcsname\relax
\expandafter\let\csname endequation*\endcsname\relax
\usepackage{amsmath}

\usepackage{graphicx}
\usepackage{rotating}
\usepackage{cite}
\usepackage{color}
\usepackage{soul}
\usepackage{siunitx}
\usepackage[colorlinks=true
,urlcolor=blue
,anchorcolor=blue
,citecolor=blue
,filecolor=blue
,linkcolor=blue
,menucolor=blue
,pagecolor=blue
,linktocpage=true
,pdfproducer=medialab
,pdfa=true
]{hyperref}
\usepackage{soul}
\usepackage[table]{xcolor}
\usepackage{multirow}

\begin{document}
	
	\setlength{\parindent}{0pt}
	
	\title[ ]{Experimental validation of particle-in-cell/Monte Carlo collisions simulations in low-pressure neon capacitively coupled plasmas}
	
\author{Chan-Won Park$^{1,2,8}$, B. Horv\'ath$^{3,4,8}$, A. Derzsi$^{3}$, J. Schulze$^{5}$, J. H. Kim$^{1}$,  Z. Donk\'o$^3$, Hyo-Chang Lee$^{6,7,*}$}

	\address{
	    $^1$ Korea Research Institute of Standards and Science, Daejeon 34113, Republic of Korea\\
	    $^2$ Department of Physics, Chungnam National University, Daejeon 34134, Republic of Korea\\
		$^3$ Institute for Solid State Physics and Optics, Wigner Research Centre for Physics, 1121 Budapest, Hungary\\
        $^4$ ELTE Eötvös Loránd University, Budapest, Hungary\\
		$^5$ Chair of Applied Electrodynamics and Plasma Technology, Faculty of Electrical Engineering and Information Sciences, Ruhr University Bochum, 44801 Bochum, Germany\\
        $^6$ School of Electronics and Information Engineering, Korea Aerospace University, Gyeonggi-do 10540, Republic of Korea\\
        $^7$ Department of Semiconductor Science, Engineering and Technology, Korea Aerospace University\\
        $^8$ First author with equal contribution.\\
        $^*$ Corresponding author
        }
        
	     \ead{plasma@kau.ac.kr}
	
	\begin{abstract}

Plasma simulations are powerful tools for understanding fundamental plasma science phenomena and for process optimization in applications. To ensure their quantitative accuracy, they must be validated against experiments. In this work, such an experimental validation is performed for a 1d3v particle-in-cell simulation complemented with the Monte Carlo treatment of collision processes of a capacitively coupled radio frequency plasma driven at 13.56 MHz and operated in neon gas. In a geometrically symmetric reactor the electron density in the discharge center and the spatio-temporal distribution of the electron impact excitation rate from the ground into the Ne 2p$_1$ state are measured by a microwave cutoff probe and phase resolved optical emission spectroscopy, respectively. The measurements are conducted for electrode gaps between 50 mm and 90 mm, neutral gas pressures between 20 mTorr and 50 mTorr, and peak-to-peak values of the driving voltage waveform between 250 V and 650 V. Simulations are performed under identical discharge conditions. In the simulations, various combinations of surface coefficients characterising the interactions of electrons and heavy particles with the anodized aluminium electrode surfaces are adopted. We find, that the simulations using a constant effective heavy particle induced secondary electron emission coefficient of 0.3 and a realistic electron-surface interaction model (which considers energy-dependent and material specific elastic and inelastic electron reflection, as well as the emission of true secondary electrons from the surface) yield results which are in good quantitative agreement with the experimental data.

	\end{abstract}

\section{Introduction}
Low-pressure capacitively coupled plasmas (CCPs) have a wide range of applications such as etching and deposition processes for semiconductor and display device fabrication \cite{Liebermann_book, Chabert_book, makabe_book, Donnelly_2013_JVST}. To optimize and control such plasma processes, fundamental insights into the plasma physics and chemistry are required to understand the effects of external control parameters such as the driving voltage/power, the reactor geometry and the  gas mixture on process relevant discharge properties \cite{Lee_2018_APR}. 

A variety of experimental and computational studies focused on different aspects of CCP operation has been conducted in the past. These include computational investigations based on fully kinetic particle-in-cell simulations complemented with the Monte Carlo treatment of collision processes (PIC/MCC) \cite{Vahedi_1993_PSST, Donko2001, Turner_2006_POP, Wilczek2016, Gudmundsson_2013, Daksha_2017_PSST, Horvath_2020_PSST,Denpoh2020,Denpoh2022,Sun2018,Gudmundsson2022,Yang_2022,Donko_2011,Wen2022} as well as hybrid fluid/kinetic simulations \cite{Wen2022,Kushner_2009,Zhang_2023} and experimental studies based on different plasma diagnostics. 

For instance, PIC/MCC simulations, that are one dimensional in space and three dimensional in velocity space (1d3v), were used to study the basic mechanisms of the formation of ion energy distribution functions (IEDFs) at the electrodes of single- and dual-frequency CCPs operated in reactive gas mixtures \cite{Georgieva_2004_PRE}. Limitations of the separate control of the ion flux and the mean ion energy at the electrodes in dual-frequency CCPs operated at strongly different frequencies were revealed based on such simulations as well, explained by the effects of frequency coupling and secondary electrons (SEs) \cite{Schulze_2009,Schulze_2011}. Both by means of experiments and simulations, voltage waveform tailoring (VWT) was demonstrated to overcome such limitations and provide an improved separate control of the mean ion energy and flux \cite{Heil_2008, Donko_2009, Schulze_2009, Korolov_2012, Lafleur_2016} as well as control over the IEDF shape \cite{Buzzi_2009, Agarwal_2005, Hartmann_2022, Hartmann_2023}. 

Hybrid simulations were used to study the effects of external control parameters on the plasma chemistry in reactive CCPs, e.g. the generation of reactive radicals \cite{Sommerer_1992_JAP,Huang_2019}. In separate experimental studies and for different discharge conditions, laser induced fluorescence \cite{Niemi_2001,Preissing_2020,Steuer_2021} and mass spectrometry \cite{Babkina_2005,Ellerweg_2010} were used to measure radical concentrations. 

The spatio-temporal dynamics of {the electron power absorption was studied both computationally and experimentally as the fundament of knowledge based plasma process development and control. Electropositive CCPs operated at low pressures and voltages were found to operate in the $\alpha$-mode \cite{BelenguerHeating,Schulze_2008,Turner_2009}, where ambipolar electron power absorption during the sheath expansion phase leads to the generation of energetic electron beams that propagate into the plasma bulk \cite{Schulze_2018}. At higher pressures/voltages or in case of the presence of an electrode with a high SE yield, such discharges were found to operate in the $\gamma$-mode, where ionization by secondary electron avalanches inside the radio frequency (RF) sheaths due to ion-induced secondary electron emission (SEE) from the electrodes dominates \cite{BelenguerHeating,Schulze_2009,Daksha_2019}. Electronegative and/or high pressure CCPs often operate in the Drift-Ambipolar (DA) mode, where strong drift and ambipolar electric fields in the plasma bulk accelerate electrons to high energies \cite{SchulzePRL_DA,Bischoff_2018,Proto_2020,Proto_2021}. At low driving frequencies, electronegative discharges can also operate in the striation mode as a consequence of the interaction of positive and negative ions with the RF electric field in the plasma bulk \cite{LiuPRL_Striation,Liu_2017}. 

Such fundamental investigations have led to the development of distinct methods of controlling the electron energy distribution function (EEDF) such as accelerating electrons in a controlled way to penetrate deeply into high aspect ratio dielectric etch profiles to compensate positive surface charges inside such trenches that would otherwise cause profile distortion and etch stops \cite{Wang_2010_JAP,  Kruger_2021,Kruger_2023,Hartmann_2022}. Multi-dimensional simulations have been used to study radial plasma non-uniformities and develop concepts for their prevention/compensation such as structured, graded conductivity and segmented electrodes \cite{Wang_2021,Park_2023,Yang_2010,Chen_2011}. Some of these concepts have been verified qualitatively based on experiments performed under different discharge conditions \cite{Schmidt_2013,Ohtsu_2016}.

These fundamental insights and control concepts were typically developed based on qualitatively correct simulation results, that can predict parameter trends correctly. Code-to-code benchmarks were performed in some cases to ensure identical computational results of several codes under identical discharge conditions \cite{Wen_2021b,Turner_2013}. While these previous works are important and of high value, simulation based plasma process development requires quantitatively accurate results in many cases. To ensure such quantitative accuracy, simulation results must be validated against experiments to make sure that they yield correct absolute values of plasma parameters such as the electron and radical densities. Code-to-code benchmark studies are not enough for that purpose, since they do not ensure realistic simulation results that agree with experiments. Experimental validation studies of plasma simulations are challenging, since they must be performed separately for different gas mixtures, reactor geometries, and discharge conditions. Moreover, well defined experimental systems must be used to ensure that their settings (reactor geometry, wall materials, etc.) agree with those assumed in the simulation. For instance, geometrically symmetric CCP reactors must be used for validating 1d3v PIC/MCC simulations, which inherently assume such reactor symmetries. Ideally, multiple plasma parameters should be measured experimentally as a function of the external control parameters and compared to the computational results. This requires using multiple diagnostics in the experiment. 

Due to these challenges such systematic experimental validation studies are rare, but represent an important topic of current research in low temperature plasma science. Such a multi-diagnostic experimental validation was recently performed for low pressure single frequency CCPs operated in argon gas in a geometrically symmetric reactor \cite{Schulenberg_2021}. The plasma density, gas temperature, IEDF, and the spatio-temporal distribution of the excitation rate of the Ar~$\rm{2p_1}$ state were measured as functions of the neutral gas pressure and the driving voltage amplitude. Good quantitative agreement with results obtained from 1d3v PIC/MCC simulations was found for a distinct set of surface coefficients used in the simulation for ion-induced SEE and effective electron reflection at the stainless steel electrodes. In a separate study, a computationally assisted diagnostics was developed to determine the ion-induced secondary electron emission coefficient (SEEC) and the electron reflection probability for various electrode materials \cite{Schulze_2022}. The basis of comparison is the IEDF obtained from measurement and PIC/MCC simulations assuming different sets of surface coefficients (which are input parameters of such simulations). This technique is conceptually similar to the computationally assisted spectroscopic technique, $\gamma$-CAST, proposed in reference \cite{Daksha_2016} for the determination of the ion-induced SEEC of electrodes exposed to CCPs based on their effects on the spatio-temporal electron impact excitation dynamics. In different mixtures of neon and oxygen gas, the spatio-temporal distributions of the Ne~$\rm{2p_1}$ excitation rate measured by PROES were compared successfully to results of 1d3v PIC/MCC simulations at different neutral gas pressures and gas mixing ratios \cite{Derzsi_2022}. Again, a strong sensitivity of the simulation results on these surface coefficients was observed and good agreement with experimental measurements was only found for distinct choices of these input parameters.      

These experimental validation studies and additional computational investigations \cite{Donko2001,Horvath_2020_PSST, Raitses_2011_IEEE, Lafleur_2013, Greb_2013_APL, Liu_2014_POP, Horvath2017, Horvath2018, Daksha_2019, Sun_2019_deltae} show that the discharge characteristics may significantly be influenced by plasma-surface interactions. Therefore, a careful selection of the most important surface processes to be included in simulations and a realistic implementation of these is important for the accuracy of the computational results. In inert gases and at low pressures of a few Pa the most important processes are electron emission and reflection due to ion- and electron-surface interactions. SEE due to excited neutrals and photons \cite{Wen_2023} as well as fast neutrals \cite{Derzsi_2015_PSST} can play a moderate role, too. The efficiency of SEE and electron reflection processes depends on the energy and the angle of incidence of the incoming particle as well as on the surface material \cite{Phelps_1999,Vaughan1,Horvath2017,Horvath2018}. The ion-induced SEE process is characterized by the corresponding SEEC, $\gamma$. At low ion impact energies ion-induced SEE is determined by the Auger effect and models exist to calculate the corresponding SEEC for metal surfaces \cite{Hagstrum,Daksha_2019}. At higher ion energies kinetic knockout effects lead to an increase of the ion-induced SEEC as a function of the ion energy \cite{Phelps_1999,Buschhaus_2022}. In the case of electron impact, various processes, including elastic and inelastic reflection and the emission of true SEs, contribute to the electron flux that leaves the surface. These processes are characterized by the coefficients $\eta_{\rm e}$, $\eta_{\rm i}$, and $\delta$ , respectively. Electron-induced SEs can impact plasma parameters such as the plasma density, the electronegativity, the metastable density in case of reactive gases and the ionization dynamics \cite{Horvath2017,Horvath2018,Wang_2021x,Wang_2021y, Horvath_2022}.

SEs induced by ions ($\gamma$-electrons) gain energy from the electric field within the sheaths and  contribute to the ionization/excitation dynamics. Therefore, the $\gamma$-electrons may even change the dominant discharge operation mode of a CCP from the $\alpha$- to the $\gamma$-mode \cite{BelenguerHeating,Horvath_2020_PSST}. In CCPs operated with $\rm{SiO_2}$ electrodes at low pressures and high voltage amplitudes, energetic ion-induced SEs generated and accelerated towards the plasma bulk at one electrode can reach the opposite electrode during the local sheath collapse and cause the emission of electron induced SEs ($\delta$-electrons) \cite{Horvath2017}. Such $\delta$-electrons can also be generated by energetic $\delta$-electrons created earlier or by bulk electrons (which are electrons generated via ionization process) that reach the electrode during sheath collapse. Depending on the time of generation, $\delta$-electrons can be accelerated towards the bulk by the residual sheath potential or by the expanding sheath. As the electron-induced SEEC can be much higher than the heavy particle induced SEECs \cite{Horvath2017,Horvath2018,Wang_2021x,Wang_2021y}, $\delta$-electrons can contribute strongly to the ionization and their presence can significantly enhance the plasma density.

The aim of this study is to validate the PIC/MCC simulation model and code for Ne gas, by conducting a systematic comparison of experimental and computational results for (i) the electron density and (ii) the spatio-temporally resolved electron impact excitation rate of the Ne~$\rm{2p_1}$ state under various plasma conditions. The effective ion-induced SEEC, $\gamma$, is determined by comparing the plasma densities obtained from the PIC/MCC simulations for various surface coefficients (used as input parameters) with a microwave cutoff probe measurement. For the electron-surface interactions, a realistic model is applied in most of the simulation cases, based on previous works \cite{Vaughan1,Horvath2017,Horvath2018,Wang_2021x,Wang_2021y, Horvath_2022}. PIC/MCC simulations and measurements are performed for different electrode gap distances, neutral gas pressures and driving voltage amplitudes of the 13.56 MHz sinusoidal driving voltage waveform to perform a quantitative and a qualitative comparison of simulation and experimental results.

The paper is structured in the following way. In Section 2, the experimental setup of the CCP with the diagnostic systems for the microwave cutoff probe measurements and for the phase resolved optical emission spectroscopy (PROES) measurements are introduced. In Section 3, some details of the PIC/MCC simulations are presented. Section 4 includes the comparison of the experimental and computational results. Finally, conclusions are drawn in section 5.

\section{Experimental setup}

\begin{figure}[h!]
	\centering
    \includegraphics[width=0.7\linewidth]{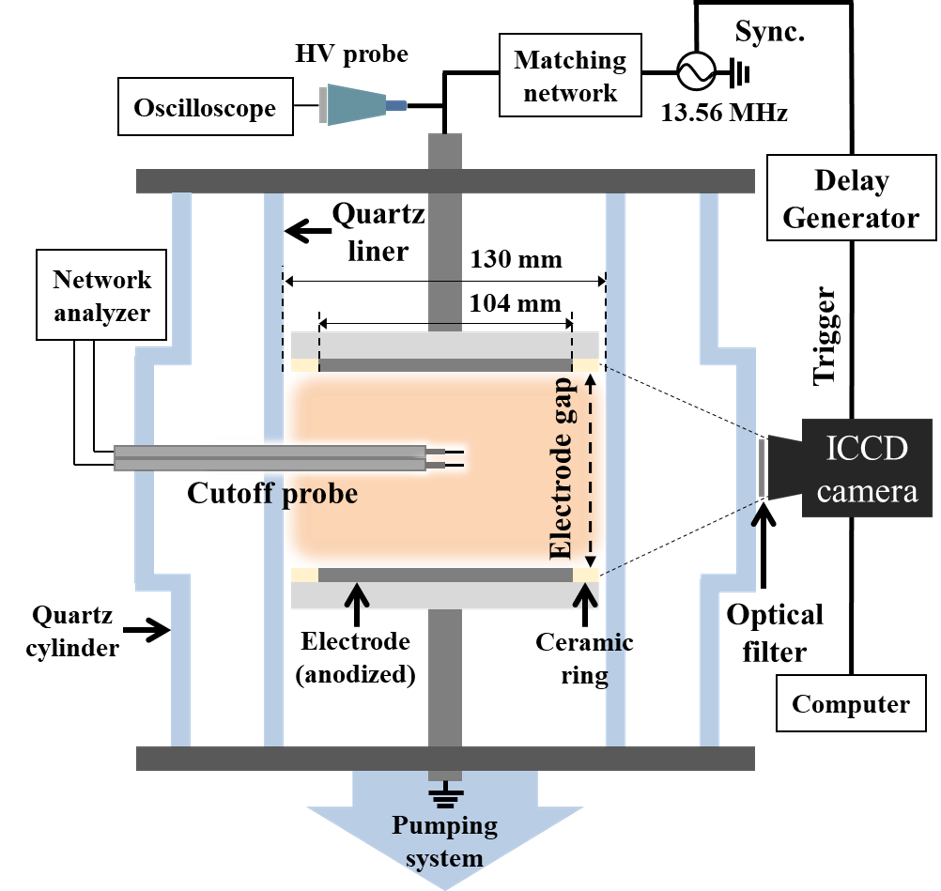}
    \caption{Schematic diagram of the experimental setup.}
    \label{fig:exp_setup}
\end{figure}

Figure~\ref{fig:exp_setup} shows a schematic diagram of the CCP reactor including the plasma diagnostics used in this work, i.e. the instrumentation for the microwave cutoff probe measurements and the PROES system. In order to realize a symmetric plasma configuration for meaningful comparisons of experimental and computational results of 1d3v PIC/MCC simulations, the plasma is generated inside a cylinder made of quartz. An additional quartz cylinder (liner) with an inner diameter of 130~mm and a height of 300~mm is used to confine the discharge in the inter-electrode space. The electrodes are made of anodized aluminium plates (Al$_2$O$_3$) with diameters of 104~mm and they are held in position by alumina rings with diameters of 127~mm. The gap distance between the two electrodes can be adjusted from 50 mm to 90 mm using stepper motors. Radio-frequency power is applied to the top electrode at a frequency of 13.56 MHz using an RF generator (RF-5S, Advanced Energy) and a matching network, while the bottom electrode is grounded. The RF voltage waveform is monitored close to the powered electrode by a high-voltage probe (P6015A, Tektronix Inc.) connected to an oscilloscope (TDS 2000B, Tektronix Inc.). A neon gas flow of 7.5 sccm is injected into the reactor via a mass flow controller and the pressure is measured using a capacitance manometer gauge calibrated with a standard pressure gauge at the Korea Research Institute of Standards and Science.

The plasma density is measured by a microwave cutoff probe \cite{Kim_2003_APL}. This  probe is inserted into the center of the plasma and is connected to a network analyzer. It consists of two antennas that transmit and receive, respectively, a frequency-swept signal. The transmitted signal passes through the plasma and is reflected at a certain frequency called the cutoff frequency, $\omega_{\rm cutoff}$, which corresponds to the electron plasma frequency, $\omega_{\rm pe}$. The electron density, $n_e$, can be obtained based on the following equation \cite{Kim_2005_Metrologia}:      

\begin{equation}
\omega_{\rm cutoff} = \omega_{\rm pe} = \sqrt{\frac{e^2 n_e}{\epsilon_0 m}}.
\end{equation}

Here, \emph{m}, \emph{e}, and $\epsilon$$_0$ are the electron mass, the elementary charge, and the permittivity of vacuum, respectively. The microwave cutoff probe directly measures the plasma density by determining this cutoff frequency with an uncertainty of less than 2\% without any specific assumptions. This accuracy holds for the plasma conditions in the presence of the probe and perturbation of the discharge by the presence of the probe itself seems to be very low due to its small size and low power input.

The PROES measurements are conducted using a fast-gateable intensified charge-coupled device (ICCD) camera (PI-MAX2, Princeton Instrument) in combination with an optical filter (585~nm band pass filter, FWHM of 16.7~nm) that measures the plasma emission at 585.5~nm resulting from electron de-excitation from the Ne~2p$_1$ state. The ICCD camera has a spatial resolution of approximately 0.5~mm and a gate time that provides a temporal resolution of 2~ns. A delay generator (DG535, Stanford Research Systems), synchronized with the RF generator, triggers the ICCD camera to acquire two-dimensional emission images of the discharge at distinct times within the RF period. These two-dimensional images are averaged in the horizontal direction and are converted to one-dimensional intensity distributions along the discharge axis in this way. The ICCD camera images are averaged for a given number of recordings for a given delay time between the camera trigger signal and gate. Then, the delay time is increased by the gate width and the discharge emission at the next phase within the RF period is measured. After scanning over one RF period in this way, the one dimensional intensity distributions  taken with different delay times provide the spatio-temporal distribution of the emission at this wavelength. 

Since cascade contributions to the population density and stepwise excitation of the Ne 2p$_1$ state are negligible, the following equation can be used to obtain the electron-impact excitation rate from the ground state into the Ne 2p$_1$ state, $\emph{E}_{0,2p_1}$, space and time resolved within the RF period from the measured emission \cite{Schulze_2007_JPD, Schulze_JPD_2010}:

\begin{equation}
E_{0,2p_1} = \frac{1}{A_{2p_1,k} n_0} \left(\frac{d \dot{n}_{ph,2p_1}(t)}{dt} + A_{2p_1} \dot{n}_{ph,2p_1}(t)  \right).
\end{equation}

Here, $n_0$ is the ground state density, $A_{2p_1,k}$ is the Einstein coefficient for the observed optical transition, $\dot{n}_{ph,2p_1}$ is the measured number of photons per unit volume and time, and $A_{2p_1}$ is the decay rate (the inverse of the lifetime) of the Ne~2p$_1$ state. The constants $A_{2p_1,k}$ and $n_0$ do not have to be known to obtain the relative changes of the electron impact excitation rate, which is sufficient to get a meaningful insight into the spatio-temporal distribution of the excitation rate into the Ne~2p$_1$ state. As the energy threshold for electron impact excitation from the ground state into the Ne 2p$_1$ state is 19 eV, the dynamics of energetic electrons above this threshold energy is revealed by measuring $\emph{E}_{0,2p_1}$ space and time resolved. 

In order to obtain experimental results as a function of external control parameters, the electrode gap (50 - 90 mm), the neutral gas pressure (20 - 50 mTorr), and the peak-to-peak value of the driving voltage waveform (250 - 650 V) are varied systematically.

\section{Simulation model}

The simulations are performed by using a 1d3v PIC/MCC simulation code. The particles traced in the simulations are electrons and Ne$^+$ ions. For the collision of electrons with atoms of the background Ne gas, elastic scattering, excitation and ionization are considered. The cross sections for these processes are taken from the Biagi-v7.1 dataset \cite{Biagi7.1}, which includes nine atomic excitation processes. The Ne~$\rm{2p_1}$ electron impact excitation process from the ground state is the one whose population dynamics is captured experimentally by the PROES measurements. For the collisions of Ne$^+$ ions with Ne atoms, isotropic and backward elastic scattering processes are considered with cross sections taken from \cite{PhepsNeonJILA}.

\begin{table}[t!]
\caption{Parameters of the realistic model of electron-surface interaction for Al$_2$O$_3$ surfaces.}
\footnotesize
\newcommand*{\TitleParbox}[1]{\parbox[c]{1.75cm}{\raggedright #1}}
\begin{tabular}{@{}lllll}
\hline \hline
\#& Parameter & Description & Value & References\\
\hline
1& $\varepsilon_{0}$  & threshold energy for electron-induced SEE & 15 eV & \cite{Gopinath}\\
2& $\varepsilon_{\rm{max,0}}$  & energy of primary electrons at the maximum emission & 600 eV & \cite{INSEPOV2010}\\
3& $\sigma_{\rm{max,0}}$  & maximum emission at normal incidence & 6 & \cite{INSEPOV2010}\\
4& $k_{\rm{s}}$  & smoothness factor of the surface & 1 & \cite{Vaughan1} \\
5& $\varepsilon_{\rm{e,0}}$  & threshold energy for elastic reflection & 0 eV & \cite{Bronstein} \\
6& $\varepsilon_{\rm{e,max}}$  & energy of primary electrons at the maximum elastic reflection & 5 eV & \cite{Barral}\\
7& $\eta_{\rm{e,max}}$  & maximum of the elastic reflection & 0.5 & \cite{Barral}\\
8& $\Delta_{\rm e}$  & control parameter for the decay of $\eta_{\rm e}$ & 5 eV & \cite{Bronstein} \\
9& $r_{\rm e}$  & portion of elastically reflected electrons & 0.03 & \cite{Gopinath}\\
10& $r_{\rm i}$  & portion of inelastically reflected electrons & 0.07 & \cite{Gopinath}\\
\hline
\label{parameters} 
\end{tabular}\\
\end{table}

\begin{figure}[t!]
	\centering
	\includegraphics[width=0.7\linewidth]{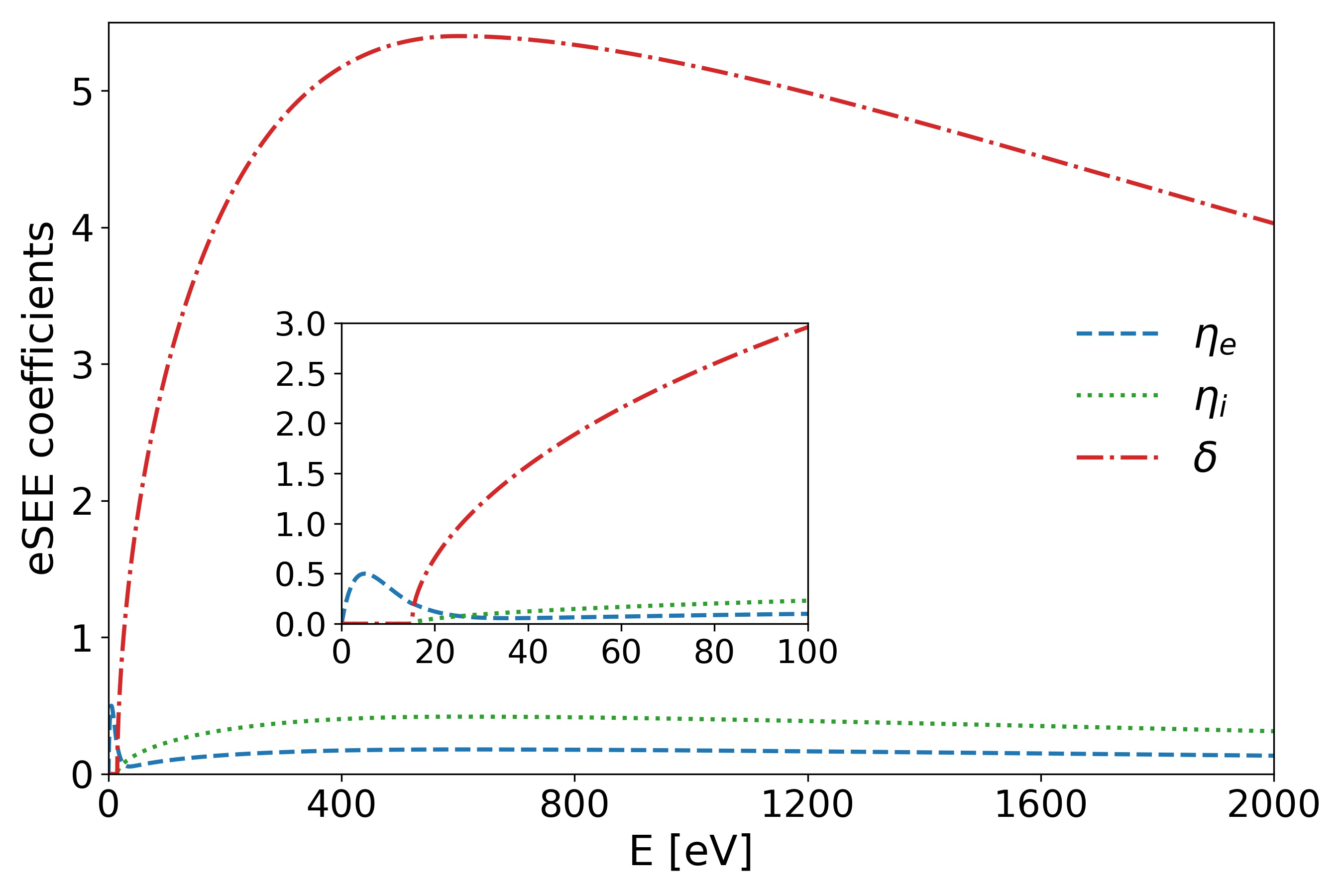}
	\caption{Energy-dependent surface coefficients for the electron-surface interactions on Al$_2$O$_3$ surfaces, for normal incidence: the total electron-induced SEEC, $\sigma$, and the partial emission coefficients of elastic electron reflection, $\eta_e$, inelastic electron backscattering, $\eta_i$, and electron-induced SEE, $\delta$. 
 }
	\label{fig:eSEE}
\end{figure}

At the surfaces, a simplified treatment of heavy particle induced SEE is considered, by assuming a constant effective SEE probability, $\gamma$, per incident Ne$^+$ ion, of which the value is varied between 0 and 0.3. For electrons, two different models are used to describe their interaction with the surfaces: 
(i) In the first approach, only elastic electron reflection is considered, by a constant $\eta_{\rm e}$ coefficient. Its value is varied between $\eta_{\rm e}=0$ (electron reflection neglected) and $\eta_{\rm e}=0.2$ \cite{Kollath1956}. (ii) In the second approach, the electron-surface interaction is described according to a realistic model, presented in \cite{Sydorenko_2006} and \cite{Horvath2017}. In this approach, the surface properties are taken into account via material specific input parameters, and the emission coefficients for the elastic reflection, $\eta_{\rm e}$, inelastic reflection, $\eta_{\rm i}$, and electron-induced SEE, $\delta$, are determined as functions of the energy and angle of incidence of primary electrons (PEs). Here, the parameters of the model are set for Al$_2$O$_3$ surfaces, based on experimental data from \cite{Gopinath, INSEPOV2010, Vaughan1, Bronstein, Barral} (see the details in table~\ref{parameters}). The total electron-induced SEEC, $\sigma=\eta_{e}+\eta_{i}+\delta$, and the partial emission coefficients for Al$_2$O$_3$ surfaces as a function of the incident electron energy at normal incidence are shown in Figure~\ref{fig:eSEE}. The simulation results obtained for the different surface coefficients are compared with experimental measurements of the plasma density and the spatio-temporally resolved electron impact excitation rate from the ground state into the Ne 2p$_1$ state to validate the simulation and find the correct values of these surface coefficients under the conditions studied experimentally.

\section{Results}

\begin{figure}[h!]
   \centering
   \includegraphics[width=0.8\linewidth]{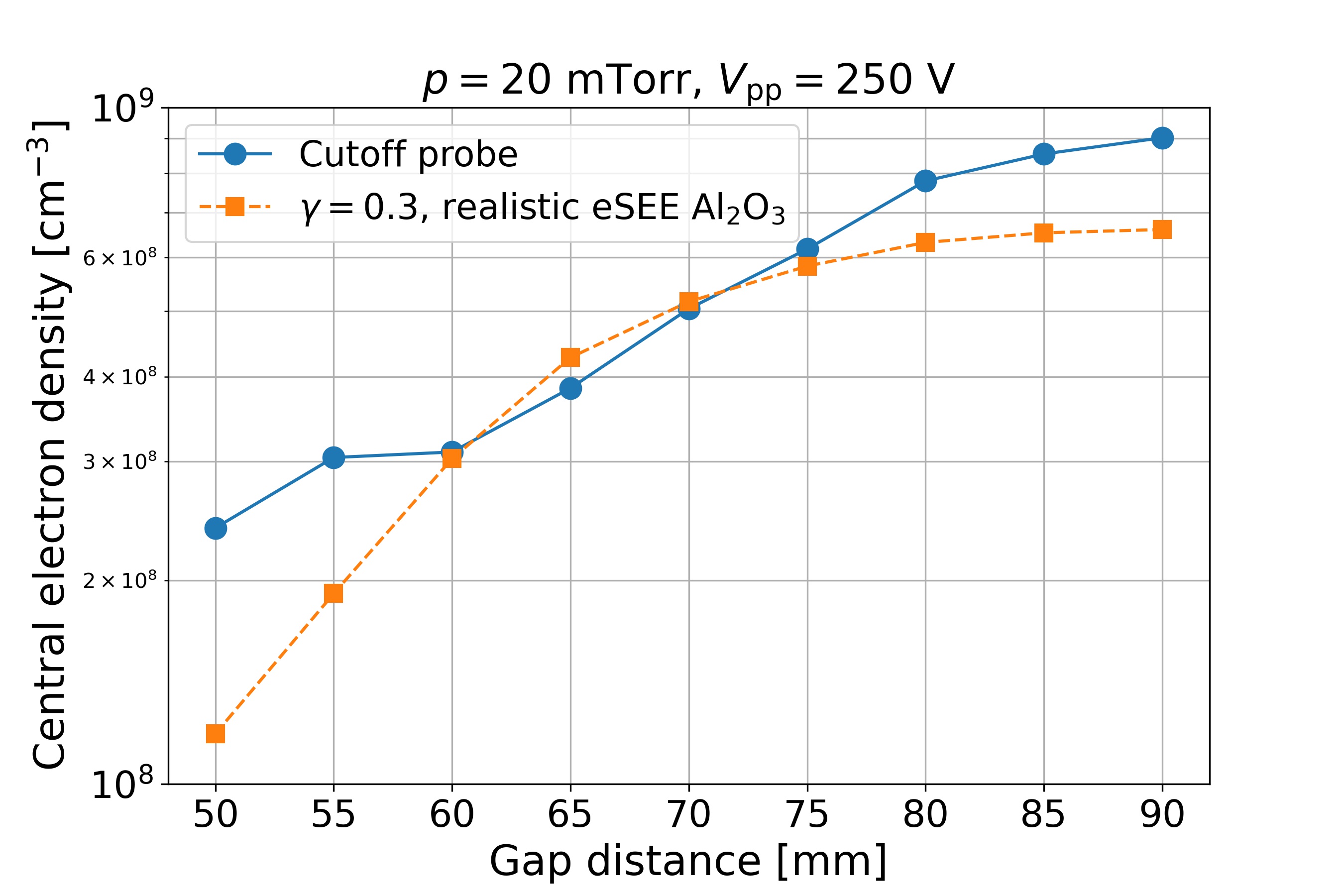}
   \caption{Central electron density measured in the center of the discharge by the cutoff probe and obtained from the PIC/MCC simulations as a function of the electrode gap at 20~mTorr and 250~V peak-to-peak voltage of the 13.56~MHz driving voltage, in the presence of Al$_2$O$_3$ electrodes. The simulation results are obtained with $\gamma = 0.3$ and the realistic electron--surface model.}
   \label{fig:cutoff vs PIC}
\end{figure}

Figure~\ref{fig:cutoff vs PIC} shows a comparison of the plasma density measured by the microwave cutoff probe in the discharge center with the corresponding plasma density obtained from the PIC/MCC simulation as a function of the electrode gap, which is varied from 50~mm to 90~mm. Ne gas is used at 20~mTorr and at a peak-to-peak driving voltage of 250~V in the presence of Al$_2$O$_3$ electrodes. In the simulation, the effective heavy particle induced and electron induced SEECs were varied systematically. The best agreement between the computational and the experimental results was found for $\gamma$ = 0.3 and by using the realistic electron-surface model. Only the simulation data for this choice of the surface coefficients are shown in figure \ref{fig:cutoff vs PIC}. The effects of using different surface coefficients on the quality of the agreement between experimental and simulation results will be discussed in more detail later. 
The plasma density increases as a function of the electrode gap in case of both the experimental and computational results. At medium values of the gap of 60~mm to 75~mm, the computed plasma density is in excellent agreement with the experimental data (with a deviation of only 3 - 10\%). For narrower and wider gap distances, the measured plasma density is higher than the one obtained from the simulation. For the lowest gap, the measured density is higher by a factor of about 2 compared to the simulation. At the largest gap the measured density is about 30\% higher. This is caused by the fact that the applicability of the 1d simulation is limited at the smallest and largest gap distances, respectively. At small electrode gaps the plasma density is very low and, thus, in the experiment a large sheath is formed at the sidewalls. This effect, which cannot be captured by the 1d simulation, in the experiment leads to additional electron power absorption and electron confinement by interactions of electrons with the large sidewall sheaths and to a constriction of the quasineutral plasma volume. Thus, more power is dissipated to electrons in a smaller volume in the experiment so that the plasma density for small electrode gaps is higher in the experiment than in the simulations, which do not include these effects. For wide electrode gaps, the applicability of the 1d simulation is limited again, since the electrode gap gets comparable to the electrode diameter in the experiment. This is no longer a 1d scenario. For instance, at the radial edge of the electrodes, where the quartz wall is in close vicinity to the electrode, a curved sheath can be formed. During the sheath expansion phase, energetic electron beams can be generated, which do not propagate in the direction perpendicular to the electrodes, but under a different angle \cite{Wang_2021}. In case of large electrode gaps, such energetic electron beams can propagate further towards the radial center of the plasma, where the plasma density is measured, and enhance the plasma density by ionization. This is qualitatively different from the scenario of smaller gaps, when such beams generated at the radial edge reach the opposite electrode quickly and get lost without propagating far in radial direction. All in all, surface effects can cause a higher plasma density in the experiment compared to the simulation at the low and the high gap region. To verify the hypotheses above, 2d simulations would be required, which are out of the scope of the current study.

\begin{figure}[ht!]
   \centering
   \includegraphics[width=0.7\linewidth]{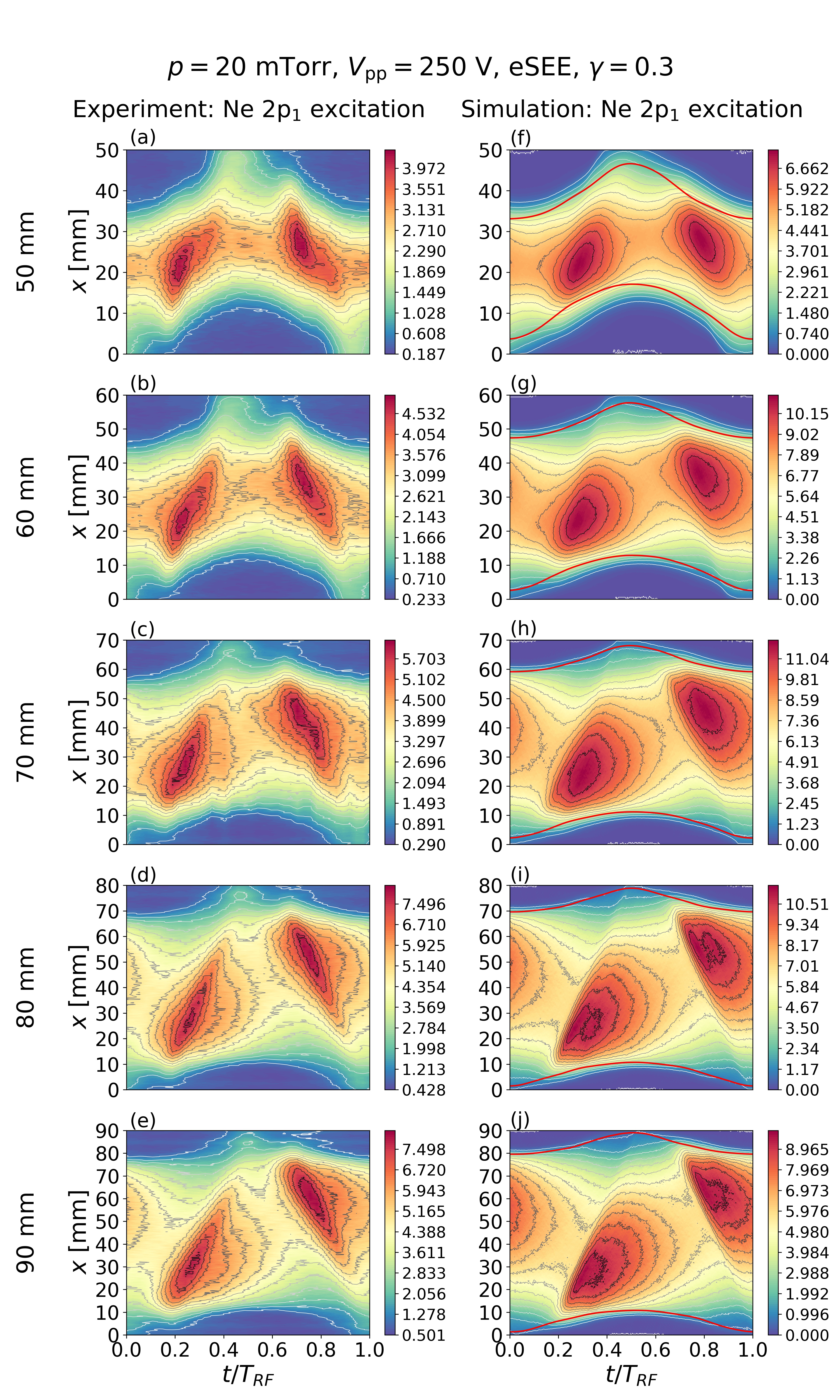}
   \caption{Spatio-temporal plots of the electron impact excitation rate from the ground state into the Ne 2p$_1$ state measured by PROES, for different electrode gaps [(a) - (e) in a.u.] and computed by the PIC/MCC simulations [(f) - (j) in units of $\rm{10^{18} ~ m^{-3} s^{-1}}$]. Discharge conditions: $f=$13.56 MHz,  $V_{\rm pp} = 250$~V, $p = 20$~mTorr. The red lines in the right column indicate the position of the sheath edge calculated according to \cite{Brinkmann2007}.}
   \label{fig:excitation rate}
\end{figure}

\begin{figure}[t!]
   \centering
   \includegraphics[width=0.7\linewidth]{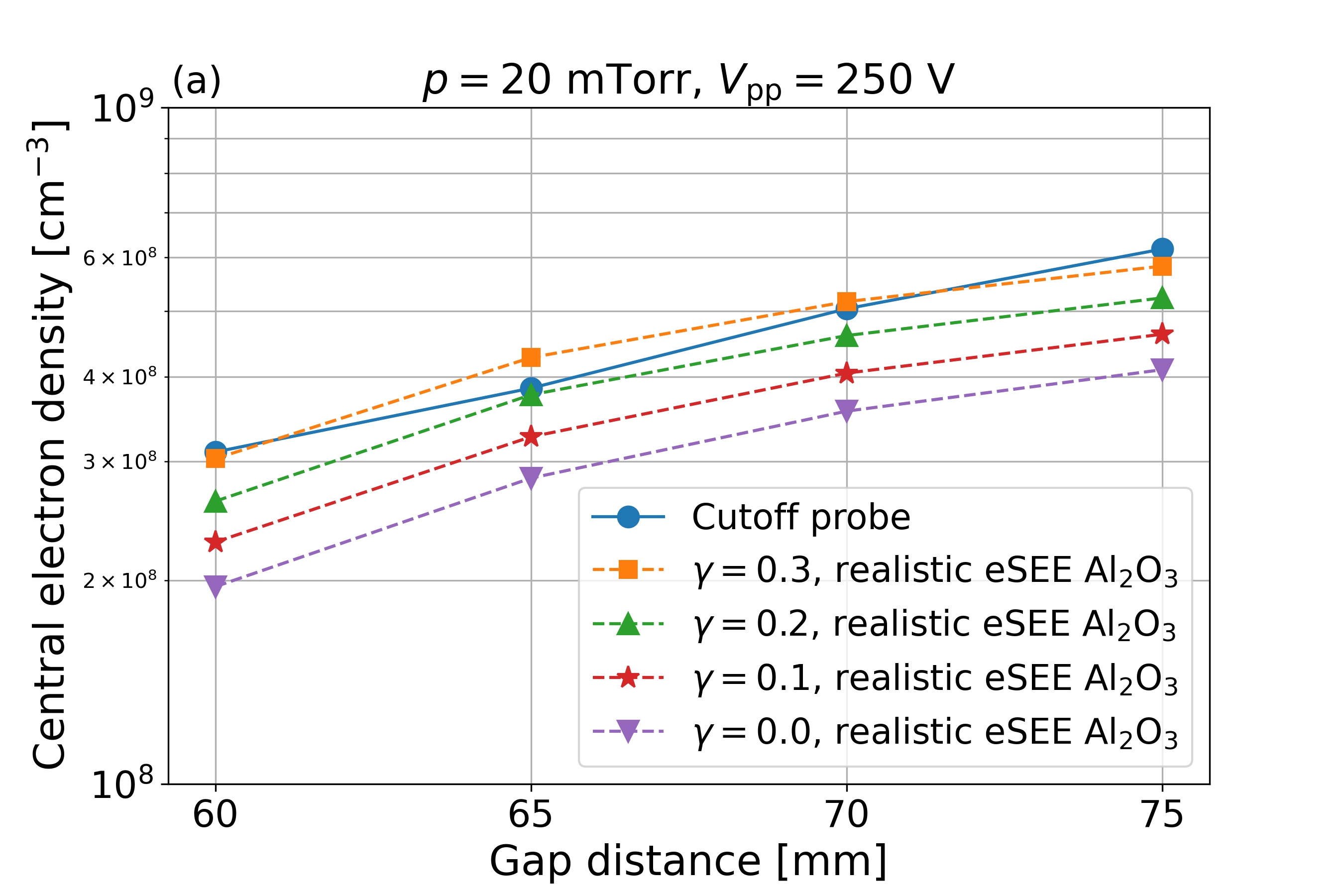}
   \includegraphics[width=0.7\linewidth]{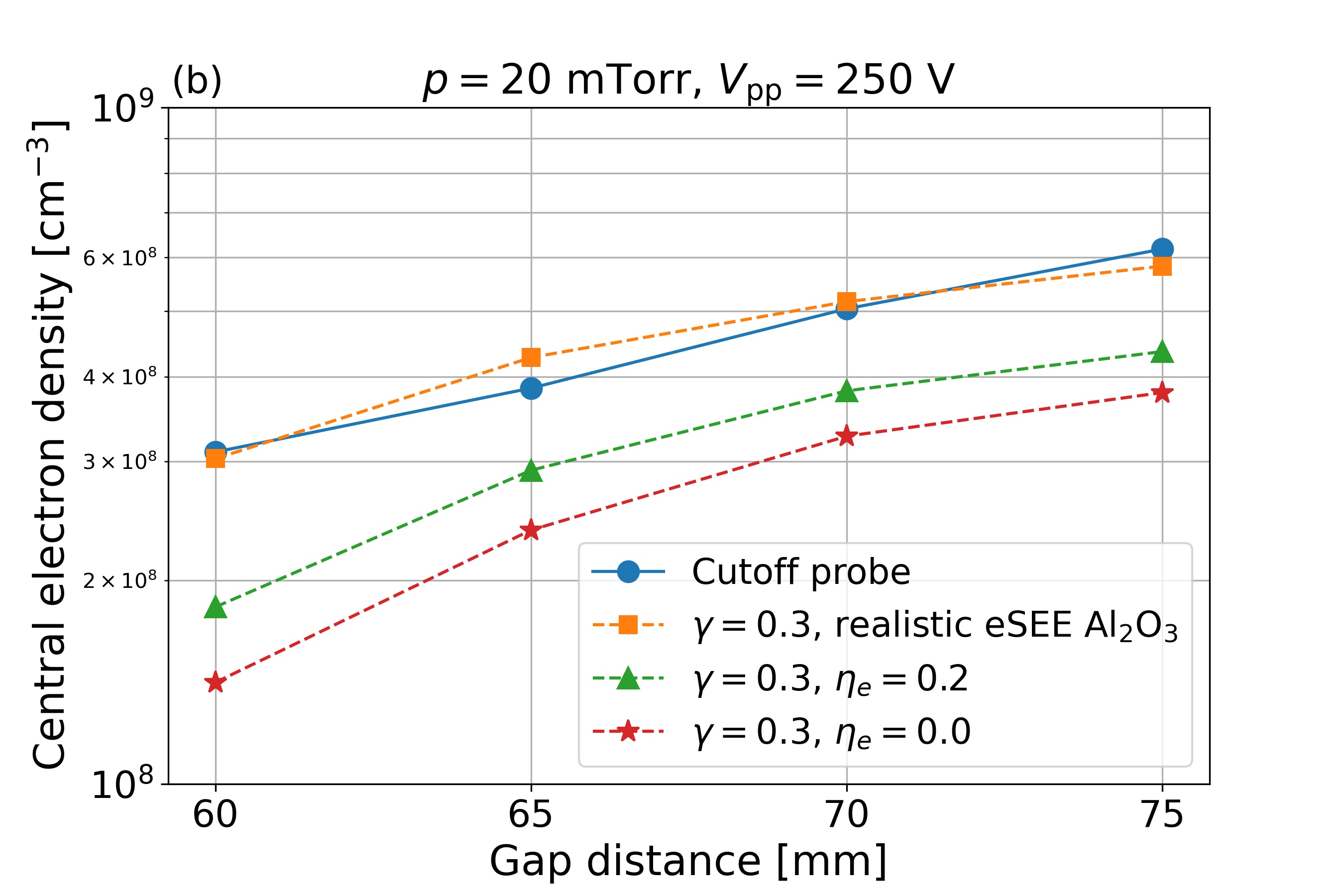}
   \caption{Plasma density obtained by microwave cutoff probe measurements and PIC/MCC simulations assuming (a) various $\gamma$-coefficients while using the realistic electron-surface model and (b) different $\eta_e$-coefficients with $\gamma$ = 0.3, as a function of the electrode gap length. Discharge conditions: neon gas, Al$_2$O$_3$ electrodes, $f=13.56$~MHz, $V_{\rm pp} = 250$~V, $p = 20$~mTorr.}
   \label{fig:SEE coefficient effect}
\end{figure}

Figure~\ref{fig:excitation rate} shows the spatio-temporal distribution of the electron impact excitation rate from the ground state into the Ne 2p$_1$ state for different electrode gaps as measured by PROES and computed by the PIC/MCC simulations for the same discharge conditions as those shown in figure \ref{fig:cutoff vs PIC}. In each panel, the vertical axis indicates the distance from the powered electrode and the horizontal axis corresponds to the time within the RF period. The left column shows the experimental results, while the right column corresponds to the simulation. The gap between the electrodes increases row by row. In the PIC/MCC simulation results, the sheath edges are shown as red lines (calculated as proposed in \cite{Brinkmann2007}), while the sheaths can be identified as blue regions of low excitation rate in the PROES results. As the electrode gap distance increases, a gradual decrease of the maximum sheath width is found in both the experimental and the computed distributions, since the ion density increases as a function of the electrode gap at fixed peak-to-peak driving voltage. For each electrode gap, the maximum sheath width is approximately the same based on the experiment and  the simulation, in accordance with the good quantitative agreement of the plasma densities (see figure~\ref{fig:cutoff vs PIC}). A general feature of these spatio-temporal plots  of the excitation rate is that energetic electron beams are generated near the sheath edge during its expansion phase and penetrate into the bulk plasma contributing to the excitation dynamics, as shown both by PROES and by PIC/MCC simulations. This is due to ambipolar/pressure heating of electrons during the sheath expansion phase \cite{Turner_1995,Schulze_2018}. The plasma is operated in the $\alpha$-mode under the conditions studied here. Overall, the spatio-temporal distributions of the excitation rate obtained by the PIC/MCC simulations are in good agreement with the PROES measurements for all electrode gaps studied. In combination with the good agreement of the plasma density, this corresponds to a successful multi-diagnostic experimental validation of the simulation for Ne gas at low pressure.

\begin{figure}[t!]
   \centering
   \includegraphics[width=0.7\linewidth]{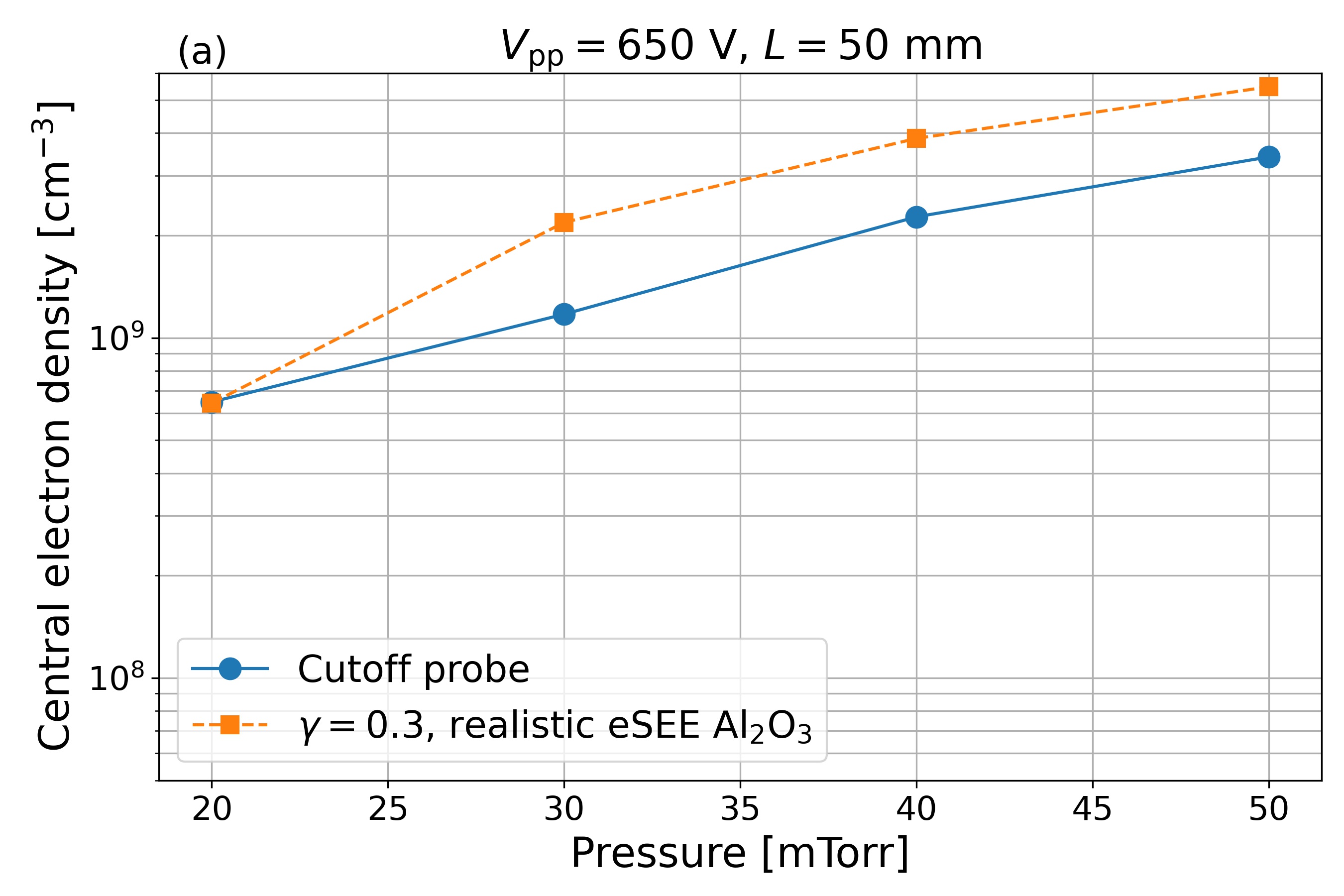}
   \includegraphics[width=0.7\linewidth]{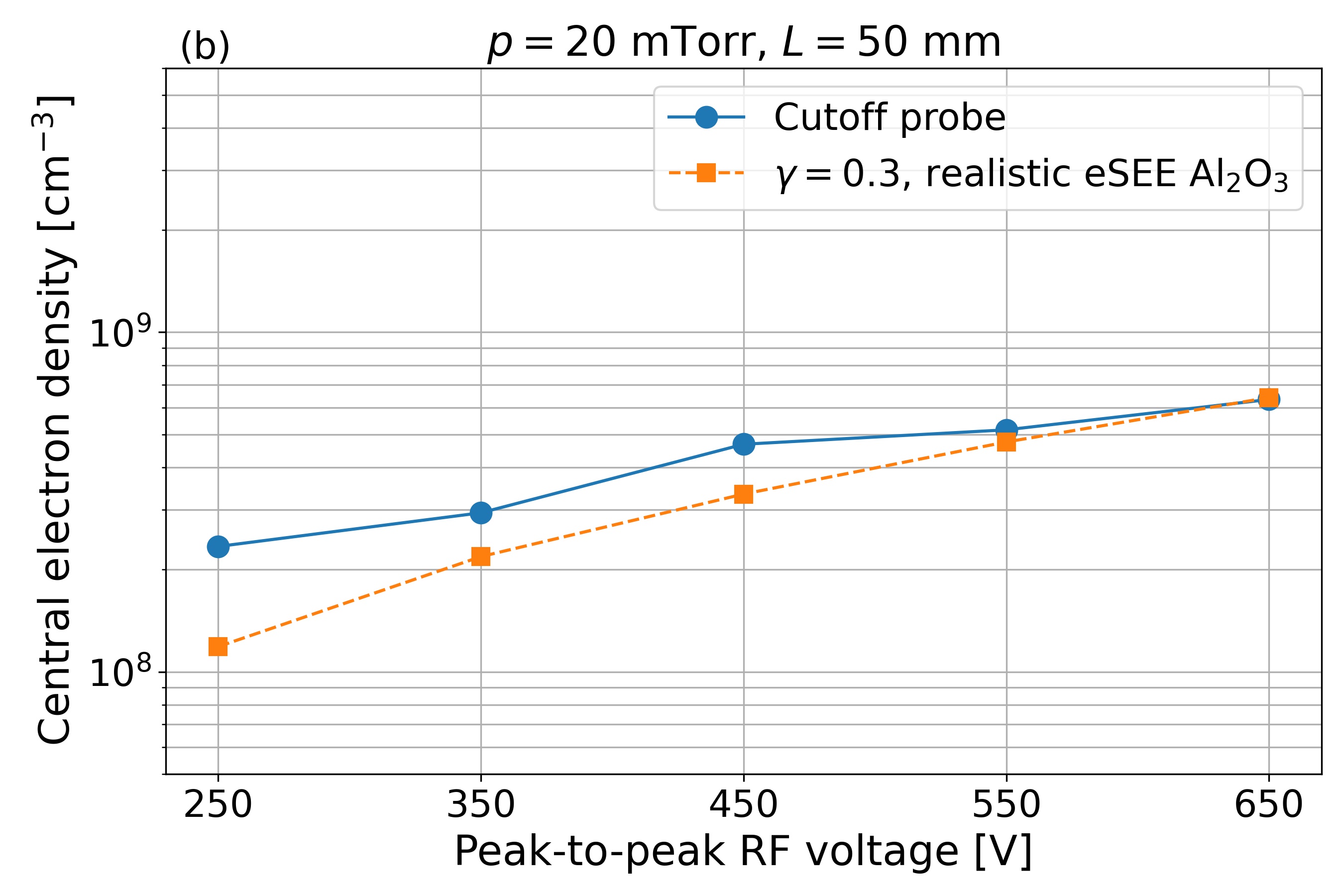}
   \caption{Plasma density obtained by microwave cutoff probe measurements and PIC/MCC simulations as a function of (a) pressure at a fixed electrode gap of 50 mm and peak-to-preak driving voltage of 650 V and (b) peak-to-peak RF voltage for a fixed electrode gap of 50 mm and pressure of 20 mTorr. $\gamma$ = 0.3 and realistic electron-surface model are used in the PIC/MCC simulation. Discharge conditions: 13.56 MHz, neon gas, Al$_2$O$_3$ electrodes.}
   \label{fig:pressure and voltage}
\end{figure}

The effects of changing the heavy particle induced SEEC, $\gamma$, and the electron surface interaction model in the simulation are illustrated by figure \ref{fig:SEE coefficient effect}, which shows the plasma density as a function of the electrode gap between 60 mm and 75 mm. As discussed before, the 1d simulation is most applicable in this range of electrode gaps and the best agreement between experimental and computational results is found for $\gamma = 0.3$ in combination with the realistic electron-surface interaction model for Al$_2$O$_3$. Decreasing $\gamma$ leads to a decrease of the plasma density obtained from the simulations (see figure \ref{fig:SEE coefficient effect}(a)) due to a lower contribution of heavy particle induced SEs to the ionization and to the emission of $\delta$-electrons. Correspondingly, a lower $\gamma$ causes worse agreement between the experimental and the simulation results. In figure \ref{fig:SEE coefficient effect}(b), the effect of keeping $\gamma=0.3$ unchanged and replacing the realistic electron-surface interaction model in the simulation by a constant elastic electron reflection probability of $\eta_e$ = 0 or $\eta_e = 0.2$ and neglecting inelastic electron reflection and electron-induced SEE is shown. Applying $\eta_e = 0.2$ results in lower plasma densities and worse agreement between the experiment and the simulation (see figure~\ref{fig:SEE coefficient effect}(b)). This happens mainly because electron-induced SEs are not included in this model and they cannot contribute to the number of electrons and the ionization dynamics. Moreover, the quality of electron confinement is reduced further by completely neglecting electron reflection by setting $\eta_e = 0$, resulting in even lower plasma densities and worse agreement with the experiment.

Figure \ref{fig:pressure and voltage}(a) shows the plasma density for a fixed electrode gap distance of 50 mm and a fixed peak-to-peak driving voltage of 650 V, as a function of the neutral gas pressure, obtained experimentally and computationally. In the simulation, the surface coefficients for which the best agreement between the experiment and the simulation was found in case of the electrode gap variation are used, i.e. $\gamma = 0.3$ and the realistic electron-surface interaction model. Reasonable qualitative agreement is found between the trends, and the densities are practically equal at the lowest pressure of 20~mTorr, while the difference is within a factor of 2 at higher pressures as well. In figure \ref{fig:pressure and voltage}(b), the plasma densities obtained from the experiment and the simulation are shown as a function of the peak-to-peak driving voltage at a fixed neutral gas pressure of 20~mTorr and electrode gap of 50~mm. Similarly to the series above, a good agreement between the experimental and the computational results is found for the same set of surface coefficients: the densities obtained from the experiment and the simulation are practically equal at the highest voltage of 650~V, and the difference is below a factor of two at lower voltages. The deviations of the simulation results from the measured density observed in figures \ref{fig:pressure and voltage}(a) and (b) at some points can be caused by the fact that in reality, the heavy particle induced SEEC changes as a function of the pressure and the voltage, since these discharge conditions influence the energy distribution of ions bombarding the electrodes, and SEE typically depends on the particle energies. This statement could be justified by including realistic SEEC for the ions in the PIC/MCC simulation, yet it remains beyond the scope of the current paper.
    
Overall, the results obtained validate the PIC/MCC simulation in neon gas under different conditions at low pressure, and they provide a good basis for future simulation based plasma process development in CCPs.

\section{Conclusions}

1d3v particle-in-cell/Monte Carlo collisions (PIC/MCC) simulations of low pressure capacitively coupled RF plasmas operated in neon at 13.56 MHz with electrodes made of Al$_2$O$_3$ were validated against experiments as a function of the electrode gap, neutral gas pressure and peak-to-peak value of the driving voltage waveform. In the frame of the experiments, the central electron density and the spatio-temporally resolved electron impact excitation rate from the ground state into the Ne 2p$_1$ state were measured by a microwave cutoff probe and phase resolved optical emission spectroscopy (PROES), respectively.

The quality of the agreement between computational and experimental results for the plasma density was found to depend on the choice of the surface coefficients for heavy particle induced secondary electron emission (SEE), $\gamma$, and electron-surface interactions. For the latter, a material specific and realistic model was found to provide the best agreement with the experimental results. In this model, elastic and inelastic electron reflection as well as electron-induced SEE were taken into account by coefficients depending on the energy and the angle of incidence of the bombarding electrons. Regarding the interaction of heavy particles with the electrodes, an effective ion-induced SEE coefficient of $\gamma = 0.3$ was found to yield good agreement with experimental results for all variations of the external control parameters performed in this work. Using a simplified electron-surface interaction model, where only constant elastic electron reflection probabilities of 0 or 0.2 were used, was found to yield remarkably lower densities compared to the experimental data.

For short electrode gaps, at which the plasma density is low and the sheaths at the sidewalls are wide, and for very large gaps, at which the electrode gap becomes comparable to the electrode diameter in the experiment, limitations of the applicability of the 1d simulations were found to reduce the quality of the agreement of the simulation with the experimental results.

Regarding the spatio-temporal distributions of the excitation rate of the Ne~2p$_1$ state, the PIC/MCC simulations provided good qualitative agreement with the PROES measurements, and the estimated width of the sheaths also matched.

Overall, within the range of electrode gaps at which the 1d simulation approach is applicable, good agreement between simulation and experimental results was found. This successful multi-diagnostic experimental validation of the 1d3v PIC/MCC simulation provides the basis of reliable plasma process development based on computational predictions. Clearly, such validation studies should be extended in the future to more complex reactive gases and gas mixtures relevant for plasma processing applications.

\ack This work was supported by the Hungarian National Research, Development and Innovation Office via grants K-134462 and FK-128924,
by the \'UNKP-22-3 New National Excellence Program of the Ministry for Culture and Innovation of Hungary from the source of the National Research, Development and Innovation Fund,
by the German Research Foundation (DFG) within the frame of the collaborative research centre SFB 1316 (project A4), 
the Material Innovation Program (Grant No. 2020M3H4A3106004) of the National Research Foundation of Korea (NRF) funded by Ministry of Science 
and ICT, the R $\&$ D Convergence Program (CRC-20-01-NFRI) of the National Research Council of Science $\&$ Technology (NST) of Republic of Korea, the Industrial Fundamental Technology Development Program of the Ministry of Trade, Industry $\&$ Energy (MOTIE) (1415187722) and the Korea Semiconductor Research Consortium (KSRC) (RS-2023-00235950).

\vspace{1cm}
	
\section*{References}

\bibliography{references}

\providecommand{\newblock}{}
\begin{thebibliography}{10}
\expandafter\ifx\csname url\endcsname\relax
  \def\url#1{{\tt #1}}\fi
\expandafter\ifx\csname urlprefix\endcsname\relax\def\urlprefix{URL }\fi
\providecommand{\eprint}[2][]{\url{#2}}

\bibitem{Liebermann_book}
 Lieberman M A and Lichtenberg A J 2005 {\em Principles of Plasma Discharges
  and Materials Processing} 2nd ed (New York: Wiley)

\bibitem{Chabert_book}
 Chabert P and Braithwaite N 2011 {\em Physics of Radio-Frequency Plasmas}
  (Cambridge: Cambridge University Press)

\bibitem{makabe_book}
 Makabe T and Petrovi\'c Z 2006 {\em Plasma Electronics: Applications in
  Microelectronic Device Fabrication} (London: Taylor and Francis).

\bibitem{Donnelly_2013_JVST}
{Donnelly V M and Kornblit A} 2013 {\em J. Vac. Sci. Technol.\/} {\bf 31}
  050825

\bibitem{Lee_2018_APR}
Lee H~C 2018 {\em Applied Physics Reviews\/} {\bf 5} 011108

\bibitem{Vahedi_1993_PSST}
{Vahedi V, Birdsall C K, Lieberman M A, DiPeso G and Ronhlien T D} 1993 {\em
  Plasma Sources Sci. Technol.\/} {\bf 2} 273

\bibitem{Donko2001}
Zoltán D 2001 {\em Phys. Rev. E\/} {\bf 64} 026401

\bibitem{Turner_2006_POP}
Turner M~M 2006 {\em Phys. Plasmas\/} {\bf 13} 033506

\bibitem{Wilczek2016}
Wilczek S, Trieschmann J, Eremin D, Brinkmann R~P, Schulze J, Schuengel E,
  Derzsi A, Korolov I, Hartmann P, Donkó Z and Mussenbrock T 2016 {\em Phys.
  Plasmas\/} {\bf 23} 063514

\bibitem{Gudmundsson_2013}
Gudmundsson J~T, Kawamura E and Lieberman M~A 2013 {\em Plasma Sources Sci.
  Technol.\/} {\bf 22} 035011

\bibitem{Daksha_2017_PSST}
{Daksha M, Derzsi A, Wilczek S, Trieschmann J, Mussenbrock T, Awakowicz P,
  Donkó Z and Schulze J} 2017 {\em Plasma Sources Sci. Technol.\/} {\bf 26}
  085006

\bibitem{Horvath_2020_PSST}
{Horváth B, Derzsi A, Schulze J, Korolov I, Hartmann P and Donkó Z} 2020 {\em
  Plasma Sources Sci. Technol.\/} {\bf 29} 055002

\bibitem{Denpoh2020}
Denpoh K 2020 {\em Japan. J. Appl. Phys.\/} {\bf 60} 016002

\bibitem{Denpoh2022}
Denpoh K and Nanbu K 2022 {\em J. Vac. Sci. Technol.\/} {\bf 40} 063007

\bibitem{Sun2018}
Sun A, Becker M~M and Loffhagen D 2018 {\em Plasma Sources Sci. Technol.\/}
  {\bf 27} 054002

\bibitem{Gudmundsson2022}
Gudmundsson J~T, Krek J, Wen D~Q, Kawamura E and Lieberman M~A 2022 {\em Plasma
  Sources Sci. Technol.\/} {\bf 30} 125011

\bibitem{Yang_2022}
Yang D, Wang H, Zheng B, Zou X, Wang X and Fu Y 2022 {\em Plasma Sources Sci.
  Technol.\/} {\bf 31} 115002

\bibitem{Donko_2011}
Donkó Z 2011 {\em Plasma Sources Sci. Technol.\/} {\bf 20} 024001

\bibitem{Wen2022}
Wen D~Q, Krek J, Gudmundsson J~T, Kawamura E, Lieberman M~A and Verboncoeur J~P
  2022 {\em IEEE Trans. Plasma Sci.\/} {\bf 50} 2548--57

\bibitem{Kushner_2009}
Kushner M~J 2009 {\em J. Phys. D: Appl. Phys.\/} {\bf 42} 194013

\bibitem{Zhang_2023}
Zhang Y~R, Huang J~W, Zhou F~J, Lu C, Sun J~Y, Su Z~X and Wang Y~N 2023 {\em
  Plasma Sources Sci. Technol.\/} {\bf 32} 054005

\bibitem{Georgieva_2004_PRE}
{Georgieva V, Bogaerts A and Gijbels R} 2004 {\em Phys. Rev. E\/} {\bf 69}
  026406

\bibitem{Schulze_2009}
Schulze J, Donk{\'{o}} Z, Luggenhölscher D and Czarnetzki U 2009 {\em Plasma
  Sources Sci. Technol.\/} {\bf 18} 034011

\bibitem{Schulze_2011}
Schulze J, Donkó Z, Schüngel E and Czarnetzki U 2011 {\em Plasma Sources Sci.
  Technol.\/} {\bf 20} 045007

\bibitem{Heil_2008}
Heil B~G, Czarnetzki U, Brinkmann R~P and Mussenbrock T 2008 {\em J. Phys. D:
  Appl. Phys.\/} {\bf 41} 165202

\bibitem{Donko_2009}
Donkó Z, Schulze J, Heil B~G and Czarnetzki U 2008 {\em J. Phys. D: Appl.
  Phys.\/} {\bf 42} 025205

\bibitem{Korolov_2012}
Korolov I, Donkó Z, Czarnetzki U and Schulze J 2012 {\em J. Phys. D: Appl.
  Phys.\/} {\bf 45} 465205

\bibitem{Lafleur_2016}
Lafleur T 2016 {\em Plasma Sources Sci. Technol.\/} {\bf 25} 013001

\bibitem{Buzzi_2009}
Buzzi F~L, Ting Y~H and Wendt A~E 2009 {\em Plasma Sources Sci. Technol.\/}
  {\bf 18} 025009

\bibitem{Agarwal_2005}
Agarwal A and Kushner M~J 2005 {\em J. Vac. Sci. Technol. A\/} {\bf 23} 1440--9

\bibitem{Hartmann_2022}
Hartmann P, Korolov I, Escandon-Lopez J, van Gennip W, Buskes K and Schulze J
  2022 {\em Plasma Sources Sci. Technol.\/} {\bf 31} 055017

\bibitem{Hartmann_2023}
Hartmann P, Korolov I, Escandon-Lopez J, van Gennip W, Buskes K and Schulze J
  2023 {\em J. Phys. D: Appl. Phys.\/} {\bf 56} 055202

\bibitem{Sommerer_1992_JAP}
Sommerer T~J and Kushner M~J 1992 {\em J. Appl. Phys.\/} {\bf 71} 1654--73

\bibitem{Huang_2019}
Huang S, Huard C, Shim S, Nam S~K, Song I~C, Lu S and Kushner M~J 2019 {\em J.
  Vac. Sci. Technol. A\/} {\bf 37} 031304

\bibitem{Niemi_2001}
Niemi K, von~der Gathen V~S and Döbele H~F 2001 {\em J. Phys. D: Appl.
  Phys.\/} {\bf 34} 2330

\bibitem{Preissing_2020}
Preissing P, Korolov I, Schulze J, Schulz-von~der Gathen V and Boeke M 2020
  {\em Plasma Sources Sci. Technol.\/} {\bf 29} 125001

\bibitem{Steuer_2021}
Steuer D, Korolov I, Chur S, Schulze J, Schulz-von~der Gathen V, Golda J and
  Boeke M 2021 {\em J. Phys. D: Appl. Phys.\/} {\bf 54} 355204

\bibitem{Babkina_2005}
Babkina T, Gans T and Czarnetzki U 2005 {\em Europhys. Lett.\/} {\bf 72}
  235--41

\bibitem{Ellerweg_2010}
Ellerweg D, Benedikt J, von Keudell A, Knake N and Schulz-von~der Gathen V 2010
  {\em New J. Phys.\/} {\bf 12} 013021

\bibitem{BelenguerHeating}
Belenguer P and Boeuf J~P 1990 {\em Phys. Rev. A\/} {\bf 41} 4447--59

\bibitem{Schulze_2008}
Schulze J, Heil B~G, Luggenhoelscher D and Czarnetzki U 2008 {\em IEEE Trans.
  Plasma Sci.\/} {\bf 36} 1400--1

\bibitem{Turner_2009}
Turner M~M 2009 {\em J. Phys. D: Appl. Phys.\/} {\bf 42} 194008

\bibitem{Schulze_2018}
Schulze J, Donko Z, Lafleur T, Wilczek S and Brinkmann R~P 2018 {\em Plasma
  Sources Sci. Technol.\/} {\bf 27} 055010

\bibitem{Daksha_2019}
Daksha M, Derzsi A, Mujahid Z, Schulenberg D, Berger B, Donk{\'{o}} Z and
  Schulze J 2019 {\em Plasma Sources Sci. Technol.\/} {\bf 28} 034002

\bibitem{SchulzePRL_DA}
Schulze J, Derzsi A, Dittmann K, Hemke T, Meichsner J and Donk\'o Z 2011 {\em
  Phys. Rev. Lett.\/} {\bf 107} 275001

\bibitem{Bischoff_2018}
Bischoff L, Hübner G, Korolov I, Donk{\'{o}} Z, Hartmann P, Gans T, Held J,
  von~der Gathen V~S, Liu Y, Mussenbrock T and Schulze J 2018 {\em Plasma
  Sources Sci. Technol.\/} {\bf 27} 125009

\bibitem{Proto_2020}
Proto A and Gudmundsson J~T 2020 {\em J. Appl. Phys.\/} {\bf 128} 113302

\bibitem{Proto_2021}
Proto A and Gudmundsson J~T 2021 {\em Plasma Sources Sci. Technol.\/} {\bf 30}
  065009

\bibitem{LiuPRL_Striation}
Liu Y~X, Sch\"ungel E, Korolov I, Donk\'o Z, Wang Y~N and Schulze J 2016 {\em
  Phys. Rev. Lett.\/} {\bf 116} 255002

\bibitem{Liu_2017}
Liu Y~X, Korolov I, Schuengel E, Wang Y~N, Donko Z and Schulze J 2017 {\em
  Plasma Sources Sci. Technol.\/} {\bf 26} 055024

\bibitem{Wang_2010_JAP}
Wang M and Kushner M~J 2010 {\em J. Appl. Phys.\/} {\bf 107} 023308

\bibitem{Kruger_2021}
Kruger F, Lee H, Nam S~K and Kushner M~J 2021 {\em Plasma Sources Sci.
  Technol.\/} {\bf 30} 085002

\bibitem{Kruger_2023}
Krueger F, Lee H, Nam S~K and Kushner M~J 2023 {\em J. Vac. Sci. Technol. A\/}
  {\bf 41} 013006

\bibitem{Wang_2021}
Wang L, Hartmann P, Donko Z, Song Y~H and Schulze J 2021 {\em J. Vac. Sci.
  Technol. A\/} {\bf 39} 063004

\bibitem{Park_2023}
Park H, Sakiyama Y and Lee H~J 2023 {\em Front. Phys.\/} {\bf 11} 1137994

\bibitem{Yang_2010}
Yang Y and Kushner M~J 2010 {\em J. Phys. D: Appl. Phys.\/} {\bf 43} 152001

\bibitem{Chen_2011}
Chen Z, Kenney J, Rauf S, Collins K, Tanaka T, Hammond N and Kudela J 2011 {\em
  IEEE Trans. Plasma Sci.\/} {\bf 39} 2526--7

\bibitem{Schmidt_2013}
Schmidt N, Schulze J, Schuengel E and Czarnetzki U 2013 {\em J. Phys. D: Appl.
  Phys.\/} {\bf 46} 505202

\bibitem{Ohtsu_2016}
Ohtsu Y, Matsumoto N, Schulze J and Schuengel E 2016 {\em Phys. Plasmas\/} {\bf
  23} 033510

\bibitem{Wen_2021b}
Wen D~Q, Krek J, Gudmundsson J~T, Kawamura E, Lieberman M~A and Verboncoeur J~P
  2021 {\em Plasma Sources Sci. Technol.\/} {\bf 30} 105009

\bibitem{Turner_2013}
Turner M~M, Derzsi A, Donko Z, Eremin D, Kelly S~J, Lafleur T and Mussenbrock T
  2013 {\em Phys. Plasmas\/} {\bf 20} 013507

\bibitem{Schulenberg_2021}
Schulenberg D~A, Korolov I, Donko Z, Derzsi A and Schulze J 2021 {\em Plasma
  Sources Sci. Technol.\/} {\bf 30} 105003

\bibitem{Schulze_2022}
Schulze C, Donkó Z and Benedikt J 2022 {\em Plasma Sources Sci. Technol.\/}
  {\bf 31} 105017

\bibitem{Daksha_2016}
Daksha M, Berger B, Sch\"uengel E, Korolov I, Derzsi A, Koepke M, Donk{\'{o}} Z
  and Schulze J 2016 {\em J. Phys. D: Appl. Phys.\/} {\bf 49} 234001

\bibitem{Derzsi_2022}
Derzsi A, Hartmann P, Vass M, Horváth B, Gyulai M, Korolov I, Schulze J and
  Donkó Z 2022 {\em Plasma Sources Sci. Technol.\/} {\bf 31} 085009

\bibitem{Raitses_2011_IEEE}
{Raitses Y, Kaganovich I D, Khrabrov A, Sydorenko D, Fisch N J, Smolyakov A}
  2011 {\em IEEE Trans. Plasma Sci.\/} {\bf 39} 995--1006

\bibitem{Lafleur_2013}
Lafleur T, Chabert P and Booth J~P 2013 {\em J. Phys. D: Appl. Phys.\/} {\bf
  46} 135201

\bibitem{Greb_2013_APL}
{Greb A, Niemi K, O'Connell D and Gans T} 2013 {\em Appl. Phys. Lett.\/} {\bf
  103} 244101

\bibitem{Liu_2014_POP}
{Liu Q, Liu Y, Samir T and Ma Z} 2014 {\em Phys. Plasmas\/} {\bf 21} 083511

\bibitem{Horvath2017}
Horváth B, Daksha M, Korolov I, Derzsi A and Schulze J 2017 {\em Plasma
  Sources Sci. Technol.\/} {\bf 26} 124001

\bibitem{Horvath2018}
Horváth B, Schulze J, Donkó Z and Derzsi A 2018 {\em J. Phys. D: Appl.
  Phys.\/} {\bf 51} 355204

\bibitem{Sun_2019_deltae}
Sun J~Y, Wen D~Q, Zhang Q~Z, Liu Y~X and Wang Y~N 2019 {\em Physics of
  Plasmas\/} {\bf 26} 063505

\bibitem{Wen_2023}
Wen D~Q, Krek J, Gudmundsson J~T, Kawamura E, Lieberman M~A, Zhang P and
  Verboncoeur J~P 2023 {\em Plasma Sources Science and Technology\/} {\bf 32}
  064001

\bibitem{Derzsi_2015_PSST}
Derzsi A, Korolov I, Schüngel E, Donk\'{o} Z and Schulze J 2015 {\em Plasma
  Sources Sci. Technol.\/} {\bf 24} 034002

\bibitem{Phelps_1999}
Phelps A~V and Petrovic Z~L 1999 {\em Plasma Sources Sci. Technol.\/} {\bf 8}
  R21

\bibitem{Vaughan1}
Vaughan J 1989 {\em IEEE Trans. Electron Devices\/} {\bf 36} 196--7

\bibitem{Hagstrum}
Hagstrum H~D 1954 {\em Phys. Rev.\/} {\bf 96} 336--65

\bibitem{Buschhaus_2022}
Buschhaus R, Prenzel M and von Keudell A 2022 {\em Plasma Sources Sci.
  Technol.\/} {\bf 31} 025017

\bibitem{Wang_2021x}
Wang L, Hartmann P, Donkó Z, Song Y~H and Schulze J 2021 {\em Plasma Sources
  Sci. Technol.\/} {\bf 30} 085011

\bibitem{Wang_2021y}
Wang L, Hartmann P, Donkó Z, Song Y~H and Schulze J 2021 {\em Plasma Sources
  Sci. Technol.\/} {\bf 30} 054001

\bibitem{Horvath_2022}
Horváth B, Donkó Z, Schulze J and Derzsi A 2022 {\em Plasma Sources Science
  and Technology\/} {\bf 31} 045025

\bibitem{Kim_2003_APL}
{Kim J-H, Seong D-J, Lim J-Y and Chung K-H} 2003 {\em Appl. Phys. Lett.\/} {\bf
  83} 4725--7

\bibitem{Kim_2005_Metrologia}
{Kim J-H, Chung K-H and Shin Y-H} 2005 {\em Appl. Phys. Lett.\/} {\bf 42} 110

\bibitem{Schulze_2007_JPD}
{Schulze J, Gans T, O'Connell D, Czarnetzki U, Ellingboe A R and Turner M M}
  2007 {\em J. Phys. D: Appl. Phys.\/} {\bf 40} 7008

\bibitem{Schulze_JPD_2010}
{Schulze J, Schüngel E, Donk{\'{o}} Z, Luggenhölscher D and Czarnetzki U}
  2010 {\em J. Phys. D: Appl. Phys.\/} {\bf 43} 124016

\bibitem{Biagi7.1}
 Biagi S F 2004 Biagi-v7.1 database \url{www.lxcat.net}, retrieved on April 8
  2019 Cross sections extracted from PROGRAM MAGBOLTZ VERSION 7.1 JUNE 2004

\bibitem{PhepsNeonJILA}
Phelps A~V personal communication

\bibitem{Gopinath}
Gopinath V~P, Verboncoeur J~P and Birdsall C~K 1998 {\em Phys. Plasmas\/} {\bf
  5} 1535--40

\bibitem{INSEPOV2010}
Insepov Z, Ivanov V and Frisch H 2010 {\em Nuclear Instruments and Methods in
  Physics Research Section B\/} {\bf 268} 3315--20

\bibitem{Bronstein}
 Bronshtein I M and Fraiman B S 1969 {\em Secondary Electron Emission} (Moscow:
  Atomizdat)

\bibitem{Barral}
Barral S, Makowski K, Peradzyński Z, Gascon N and Dudeck M 2003 {\em Phys.
  Plasmas\/} {\bf 10} 4137--52

\bibitem{Kollath1956}
 Kollath R 1956 {\em Encyclopedia of Physics} ed S Fl\"ugge vol 21 (Berlin:
  Springer) p 264

\bibitem{Sydorenko_2006}
 D. Sydorenko 2006, {\em Particle-in-cell simulations of electron dynamics in
  low pressure discharges with magnetic fields.} PhD Thesis, University of
  Saskatchewan, Saskatoon, Canada.

\bibitem{Brinkmann2007}
Brinkmann R~P 2007 {\em Journal of Appl. Phys.\/} {\bf 102} 093303

\bibitem{Turner_1995}
Turner M~M 1995 {\em Phys. Rev. Lett.\/} {\bf 75} 1312--5

\end{thebibliography}
\bibliographystyle{iopart-num}
	
\end{document}